\documentclass[aps,twocolumn,amsmath,amssymb,showpacs,prl]{revtex4}
\usepackage{epsf}

\newcommand{\bk}{{\boldsymbol k}}

\newcommand{\bR}{{\boldsymbol R}}

\newcommand{\bp}{{\boldsymbol p}}
\newcommand{\bz}{{\boldsymbol z}}
\newcommand{\bd}{{\boldsymbol d}}

\begin{document}

\title{Excess current in superconducting Sr$_2$RuO$_4$}  
\author{ F.\ Laube$^{1}$, G.\ Goll$^{1}$, 
M.\ Eschrig$^{2}$, M. Fogelstr{\"o}m$^{3}$, 
and Ralph Werner$^{4}$ }   
\affiliation{
$^1$Physikalisches Institut, Universit\"at Karlsruhe, D-76128
 Karlsruhe, Germany\\
$^2$Institut f\"ur Theoretische Festk\"orperphysik, Universit\"at
 Karlsruhe, D-76128 Karlsruhe, Germany\\
$^3$Applied Quantum Physics and Complex Systems, MC2, Chalmers, S-41296 G\"oteborg, Sweden\\
$^4$Institut f\"ur Theorie der Kondensierten 
\mbox{Materie, Universit\"at} Karlsruhe, D-76128 Karlsruhe, Germany } 
\date{\today}
\begin{abstract} 
We present results from point-contact measurements on Sr$_2$RuO$_4$
that show a linear dependence of the excess current as a function 
of temperature and applied magnetic fields over a surprisingly wide
range of the phase diagram. 
We propose an explanation of this finding in terms of a $p$-wave 
triplet-pairing 
state with coupling to a low-energy fluctuation mode. Within this
model we obtain a quantitative description of the temperature dependence
of the excess current.
The impact 
of surface effects on order parameter and excess current is addressed.  
\end{abstract}
\pacs{74.70.Pq,73.40.Jn,74.20.Rp,74.20.-z}
\maketitle

{\it Introduction.}---The discovery of superconductivity below
$T_c\sim 1.5$ K in Sr$_2$RuO$_4$ has quickly triggered a large amount
of interest because of the unconventional properties \cite{MRS01} and
the initially proposed analogy \cite{RS95} to $^3$He. 
The enhanced specific heat, magnetic susceptibility, and electronic
mass indicate the presence of significant correlations
\cite{NMF+98,BJM+00,DLS+00,BMJ+02}. 
For a more detailed overview see Refs.\ \onlinecite{MRS01} and
\onlinecite{BMJ+02}.  
The exact symmetry of the superconducting order parameter (OP)
\cite{HS98,SAF+99,GB00,EFF01,KS02,Wern02e} and notably the pairing 
mechanism \cite{Bask96,SBB+99,MS99,EMJB02,OO02b} are still
controversial. Andreev spectra are sensitive to the order-parameter
symmetry \cite{HS98,Esch00} and are thus an adequate experimental probe to
yield clarifying information. The shape of the spectra previously
obtained from point-contact measurements in superconducting
Sr$_2$RuO$_4$ where satisfactorily reproduced by an analysis of a
$p$-wave pairing state with OP $\bd(\bk)=\hat{\bz}(k_x\pm
i k_y)$ \cite{LGL+00}.

\begin{figure}
   \epsfxsize=0.48\textwidth
   \centerline{\epsffile{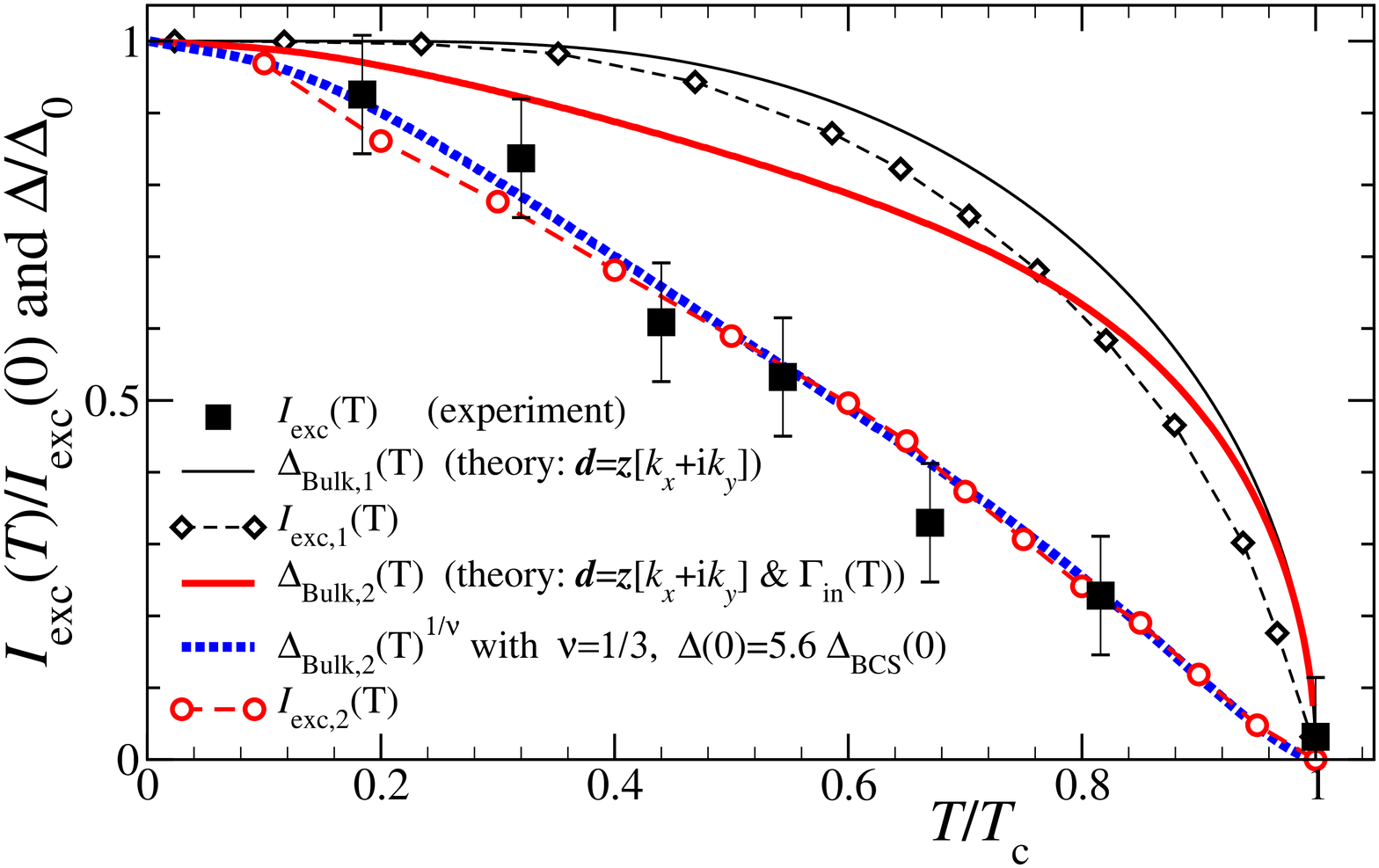}}
   \caption{\label{IexcofT}
   Temperature dependence of the normalized excess current across
   a point contact in Sr$_2$RuO$_4$. Experimental results (squares)
   are taken from Ref.\ \protect\onlinecite{LGL+00}. Open symbols show
   the results of our calculation for the excess current from a
   $p$-wave analysis, without (diamonds) and with (circles)
   the effects of an inelastic 
   scattering channel $\Gamma_{\rm in}(T)$.
   The dashed thick curve illustrates the scaling
   relation $I_{\rm exc,2}(T)\propto \Delta_{\rm Bulk,2}(T)^{\frac{1}{\nu}}$
   for the inelastic scattering model. 
   } 
   \end{figure}

It has been shown that the excess current in $s$-wave superconductors
is proportional to the superconducting gap \cite{BTK82} and
consequently contains further information on the superconducting
state. The present paper discusses excess-current measurements in
Sr$_2$RuO$_4$ in greater detail.
In particular, we present experimental results for the excess current in
applied magnetic fields and find a systematic linear behavior as a
function of field over a surprisingly wide range. This finding is
compared with the experimental excess-current data from Ref.\
\onlinecite{LGL+00} (squares in Fig.\ \ref{IexcofT}), which exhibit a
strikingly linear temperature dependence as well. We show that these
two findings imply a well defined functional relationship between the
excess current and the OP, and discuss the resulting implications in
the framework of the $p$-wave picture \cite{HS98,SAF+99,LGL+00} 
extended to include effects of low-energy fluctuations
\cite{Millis88,Wern03b}. 
Our measurements also suggest the presence of a normal-conducting
surface layer in Sr$_2$RuO$_4$.
We model such a layer by an enhanced scattering rate near the
surface and obtain qualitative agreement with the experimental
point-contact spectra within the $p$-wave picture.

{\it Experiment.}---Our measurements were performed on two single
crystals grown both by a floating zone technique in  different groups.
$T_c$ was obtained via bulk resistivity measurements. One crystal,
labeled \#5 (Ref.\ \onlinecite{LCMS92}) shows a midpoint transition
temperature $T_c^{50 \%}=1.02$~K with a transition width $\Delta
T_c^{90\%-10\%}=0.035$~K and the other, \#C85B5 (Ref.\
\onlinecite{MMF00}) has $T_c^{50 \%}=1.54$~K and $\Delta
T_c^{90\%-10\%}=0.15$~K. Heterocontacts between superconducting
Sr$_2$RuO$_4$  and a sharpened Pt needle as a counter-electrode were
realized inside the mixing chamber of a $^3$He/$^4$He dilution
refrigerator. The differential resistance $dV/dI$ vs.\ $V$ was
recorded by a standard lock-in technique. The differential
conductance, $dI/dV$, is obtained by numerical inversion of the
measured $dV/dI$ data. 
Measurements were performed in different configurations with respect to
the predominant current injection relative to the crystallographic axis
of Sr$_2$RuO$_4$, the applied magnetic field, and the surface
treatment \cite{Laub02}. Here we focus on results obtained for
contacts with $j||ab$ and $H||c$ within an accuracy of about 5-10$^\circ$.

\begin{figure}
   \epsfclipon
   \epsfxsize=0.48\textwidth
   \centerline{\epsffile{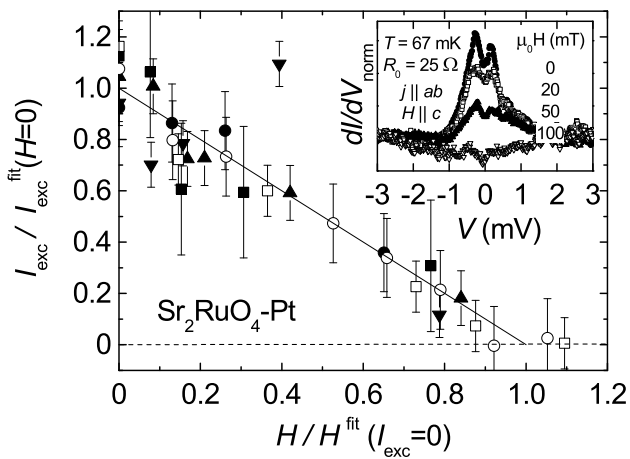}}
   \epsfclipoff
   \caption{\label{IexcofB}
   Field dependence of the normalized excess current 
   across several point contacts
   in Sr$_2$RuO$_4$. The magnetic field $H$ is aligned almost parallel
   to the $c$-axis, and the current accross the point contact is
   applied in the $ab$-plane. Each symbol represents one point contact
   on one of the two studied samples. The full line is a guide to the
   eye. For explanation of $I_{\rm exc}^{\rm fit}(H=0)$ and $H^{\rm
   fit}(I_{\rm exc}=0)$ see text. The inset shows for one point
   contact typical $dI/dV$ curves from which the excess current was
   determined as a function of magnetic field.}
   \end{figure}

{\it Linear field dependence.}---We focus on high
transmission contacts, which exhibit a double-minimum structure in the
differential resistance, i.e., a double-maximum in the differential
conductance $dI/dV$ vs.\ $V$ (see inset of Fig.\ \ref{IexcofB}). In
this metallic regime, in contrast to the tunneling limit, excess
current via the mechanism of Andreev reflection occurs. Figure
\ref{IexcofB} shows the observed magnetic field dependence of the
normalized excess current (symbols) across several point contacts in
Sr$_2$RuO$_4$ measured at temperatures $0.04<T/T_c<0.20$, together
with a linear guide to the eye (line). Each symbol represents one
point contact either on sample \#5 (open symbols) or \#C85B5 (filled
symbols). The excess current was determined by numerical integration
of $dI/dV$ vs. $V$ after subtraction of a normal-conducting
background. The normalization values $I_{\rm exc}^{\rm fit}(H=0)$ and
$H^{\rm fit}(I_{\rm exc}=0)$ have been determined from linear
regression of the $I_{\rm exc}(H)$ vs.\ $H$ data for each point
contact separately. The inset shows typical $dI/dV$ curves from which the
excess current was derived.

The data in Fig.\ \ref{IexcofB} clearly are consistent with a linear
dependence of the excess current on the applied magnetic
field. Note that the result is quite universal since it is found in
both samples in spite of their different value of $T_c$. As will be
outlined now, the observed equivalence of the linearity in the field
and temperature dependence implies a well defined functional
dependence of the excess current $I_{\rm exc}$ on the superconducting
gap $\Delta$.

{\it Scaling relation for $I_{\rm exc}$.}---Consider the modulus of the
superconducting OP near the field dependent critical
temperature $T_{\rm c}(H)$ as a function of the reduced temperature
$t(H) = 1 - \frac{T}{T_{\rm c}(H)}$ at a given magnetic field: 
\begin{equation}\label{Doft}
\Delta_H(T)\big|_{t \ll 1} = A_H \ t^\nu\,.
\end{equation}
The mean-field exponent ($p$-wave approach) is $\nu=1/2$ while in the
recently introduced third-order transition picture \cite{Wern03c} one
has $\nu=1$. The proportionality factor $A_H$ depends on the magnetic
field, $H$. In order to find the resulting field dependence of the OP
modulus consider the phenomenological interpolation formula  
\begin{equation}\label{HcofT}
\frac{H_{\rm c2}(T)}{H_{\rm c2}(T=0)} =  1 - 
                    \left(\frac{T}{T_c(H=0)}\right)^2
\end{equation}
determining 
the upper critical magnetic field,
$H_{\rm c2}$, as a function of temperature. Eq.\ (\ref{HcofT})  with
$\mu_0 H_{\rm c2}(T=0)=1.5$ T and $T_c(H=0)=1.5$ K reproduces the
experimental data \cite{MMM99} satisfactorily. The inverse of Eq.\
(\ref{HcofT}) determines $T_c(H)$. Defining the reduced field $h(T)
= 1 - \frac{H}{H_{\rm c2}(T)}$ at a given temperature and
expanding Eq.\ (\ref{HcofT}) for $t\ll 1$ one finds the relation   
\begin{equation}\label{tofH}
t(H)\ \frac{T_{\rm c}^2(H)}{T_{\rm c}^2(0)}
\approx \frac{1}{2}
\frac{H_{\rm c2}(T)}{H_{\rm c2}(0)}\ h(T)\,
\end{equation}
between the reduced temperature and the reduced field. Consequently
the reduced field dependence of the gap at a given temperature is 
\begin{equation}\label{Dofh}
\Delta_T(H)\big|_{h \ll 1} = A_H\ 
       \frac{[H_{\rm c2}(T)/2]^\nu}{[H_{\rm c2}(0) - H]^\nu}\ 
       h^\nu\,.
\end{equation}

For $A_H = \mbox{constant}$ the pre-factor of the right hand side of
Eq.\ (\ref{Dofh}) implies anomalies for low temperatures near the
critical field, notably $\lim_{h\to 0}[\lim_{T\to 0} \Delta_T(H)] \neq
0$. Since $\Delta_T(H)$ and $I_{\rm exc}$ are closely related (Ref.\
\onlinecite{BTK82} and below), the observed linearity in Fig.\
\ref{IexcofB} requires that the divergence is compensated by the
pre-factor \cite{Wern03c} through $A_H \sim [H_{\rm c2}(0) - H]^\nu$ 
and hence 
\begin{equation}\label{IofD}
I_{\rm exc} = \mbox{constant}\times \Delta^{1/\nu}\,.
\end{equation}
Eq.\ (\ref{IofD}) marks a central result as it imposes a necessary
condition on any theoretical approach to the superconductivity in
Sr$_2$RuO$_4$ in order to satisfy Eqs.\ (\ref{Doft}) and (\ref{Dofh})
together with the experimental observations in Figs.\ \ref{IexcofT}
and \ref{IexcofB}. In the light of this scaling relation we discuss
the widely accepted triplet $p$-wave pairing scenario, which we extend
to include effects of low-lying bosonic fluctuations.

{\it $P$-wave scenario.}---The $p$-wave analysis of the point contact
spectra applied in Ref.\ \onlinecite{LGL+00} yields the BCS
temperature dependence for the
superconducting gap in Sr$_2$RuO$_4$, shown as $\Delta_{\rm Bulk,1}$ 
(thin full line) in Fig.\ \ref{IexcofT}. 
The resulting temperature dependence of the excess
current, determined from the calculated conductance for Andreev type
spectra, in the framework of the $p$-wave analysis is also
shown in Fig.\ \ref{IexcofT} as $I_{\rm exc,1}$ (diamonds).
The calculations are performed for a mean free path of
15 coherence lengths ($\xi_0=v_f/2\pi T_c$) and for a diffusely
scattering surface modelled as in Ref.\ \onlinecite{LGL+00}.
It is clear that this model is insufficient to describe the experimental data.
Nevertheless, it is interesting to note that
unlike in the
$s$-wave case in unconventional superconductors the excess current is
not necessarily proportional to the OP; we find near $T_c$
a temperature variation of the excess current linear in $t$, in contrast
to the $t^{1/2}$ variation of the OP.
This is because impurities and disorder strongly affect the surface
properties of unconventional superconductors \cite{Kopn86}.

{\it Pair-breaking by low-frequency bosonic fluctuations.}---As seen above, 
the $p$-wave scenario alone does not account for the observed temperature
dependence of the experimentally obtained $I_{\rm exc}(T)$. This
is true also for the overall magnitude of the bulk gap,
$\Delta(0)=1.1 {\rm meV}=6 \times 1.76 k_{\rm B} T_{\rm c}$, extracted from
tunneling spectra \cite{LGL+00}. To reconcile the measured $\Delta(0)$
and $I_{\rm exc}(T)$ with a $p$-wave OP we consider an additional 
pair-breaking channel. It was shown by Millis {\it et. al.} \cite{Millis88}
that a low-frequency bosonic mode at a characteristic
frequency $\omega_p$ 
described by an Einstein spectrum
$A_p(\omega)=\frac{\pi}{2}J_p\omega_p \delta(\omega-\omega_p)$ 
leads
to a {\em temperature dependent} pair-breaking parameter 
\begin{equation}
\Gamma_{\rm in}(T)=\frac{(1-g)}{4}J_p\omega_p \coth(\frac{\omega_p}{2T})\;,
\end{equation}
where $g$ is the coupling-constant appearing in the gap equation.
The assumptions are
that $\omega_p < T_{\rm c} \ll \omega_{\rm E}$,  
$\omega_{\rm E}$ being the frequency of the pairing mode,
and that $A_p(\omega)$ is
unaffected by the transition into the superconducting state. 
We performed calculations using the quasiclassical Green's functions technique
and included the pair-breaking parameter as a 
self-energy within a self-consistent Born approximation, i.e.\
$\hat \Sigma_{\rm in}(\bR,\varepsilon,T)=\Gamma_{\rm in}(T)\,
\langle \hat g(\bp_f,\bR,\varepsilon)\rangle_{\bp_f}$, 
where $\epsilon $ is the energy of the quasiparticles, 
$\bR $ the position with respect to the interface, and
$\bp_f$ the Fermi momentum; the $\bp_f$-average is over the Fermi surface.
The Green's function $\hat g(\bp_f,\bR,\varepsilon)$ is a functional of
the self-energy $\hat \Sigma_{\rm in}(\bR,\varepsilon,T)$ in the usual way.
The order parameter profile $\Delta (\bR,T)$ near the interface was then
obtained by iterating the weak-coupling gap equation and 
$\hat \Sigma_{\rm in}(\bR,\varepsilon,T)$
until convergence.

For the excess current this model gives an excellent agreement with 
experimental data, as shown by $I_{\rm exc,2}$ (circles) in Fig.\ \ref{IexcofT}
for $\omega_p=0.5 T_{\rm c}$
and $\frac{(1-g)}{4}J_p=2 \pi \times 0.25$. 
The almost linear temperature dependence over the whole temperature range is
reproduced within our model, and furthermore,
as shown as the dashed thick line in Fig.\ \ref{IexcofT},
the above introduced scaling relation between the calculated $I_{\rm exc,2}(T)$ 
and the theoretically obtained
order parameter $\Delta_{\rm Bulk,2} (T)$ is fulfilled to remarkable accuracy 
with the scaling exponent $\nu = 1/3$.

Another effect of $\Gamma_{\rm in}(T)$ is that the enhanced scattering at
higher temperatures 
reduces the observed $T_{\rm c}$ substantially from its ideal value 
while the gap at $T\rightarrow 0$ is
much less affected, giving $\Delta(0)/k_{\rm B} T_{\rm c}$-ratios
much larger than the BCS-ratio 1.76. 
Our calculations give the correct
absolute magnitude, $\Delta(0)=5.6 \Delta_{\rm BCS}(0)$. 
Notably, also the functional form of $\Delta(T)/\Delta(0)$ 
is modified compared to the pure $p$-wave case
(see $\Delta_{\rm Bulk,2}$, thick line in Fig.\ \ref{IexcofT}).
The conductances
calculated with the present model have the same qualitative features,
both for the Andreev and the tunnel limit, as those displayed in
Ref. \cite{LGL+00}, and can still account for the measured data. 

{\it Normal-state surface layer.}---In order to 
obtain more detailed insight about the nature of the pairing state in
Sr$_2$RuO$_4$ it would be
instructive to quantify empirically the field dependence of $A_H$ in
Eqs.\ (\ref{Doft}) and (\ref{Dofh}). Unfortunately, obtaining data
from the necessary temperature scans at different fields for a given
point contact is difficult because of the sensitivity of the large
background resistivity \cite{LGL+00} in the $dV/dI$ data to very small
changes in the configuration. A possible reason for the presence of a
large background can be found in a normal-state surface layer due to
surface reconstructions \cite{SDL+01} that leads to an additive
resistivity in the point contact as 
$\sigma^{-1}_{\rm measured} = R_{\rm N} + \sigma^{-1}_{\rm N-S}$. 
Here $\sigma^{-1}_{\rm N-S}$ is the conductivity of the
nor\-mal-super\-con\-duc\-tor interface, $R_{\rm N}$ is the normal
layer resistivity, and typically $\sigma^{-1}_{\rm N-S} / R_{\rm N}
\sim 10$\%. Note that the thickness of the
normal-state surface layer appears to be independent of the sample
quality since the observed values of $0.5\ \Omega \le R_{\rm N} \le
25\ \Omega$ vary from point contact to point contact but are in the
same range for both samples \cite{Laub02}.

   \begin{figure}
   \epsfxsize=0.46\textwidth
   \centerline{\epsffile{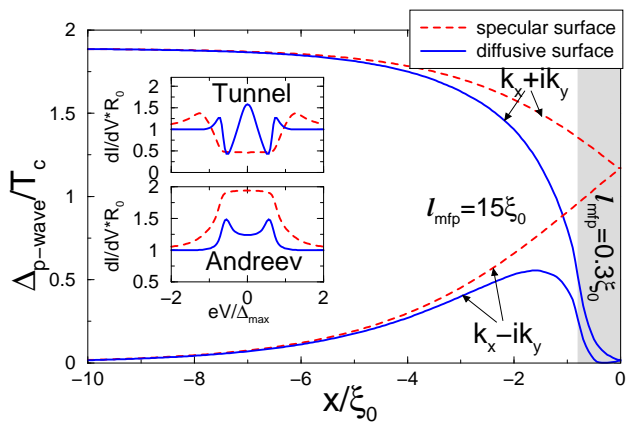}}
   \caption{\label{normallayer}
   Creation of a normal-conducting surface layer in a $p$-wave
   superconductor due to an increased scattering rate near the surface.
   For comparison, as dashed lines are shown the curves assuming a clean
   surface. 
   The two OP components are the bulk
   $k_x+ik_y$ order parameter, and the subdominant $k_x-ik_y$ order
   parameter which is stabilized only within a few coherence lengths
   ($\xi_0=v_f/2\pi T_c$) near the surface.
   Both OP components are suppressed in a layer with 
   increased scattering, leading effectively to a normal-conducting 
   surface layer. The calculations are for $T=0.05 T_c$. 
   The insets show the corresponding point-contact spectra (bottom) and
   tunneling spectra (top).
   }
   \end{figure}

Such a normal-state surface layer has a natural explanation in
a $p$-wave triplet scenario because the $p$-wave OP is very sensitive
to scattering. We assume a region near the interface in which
scattering is enhanced. 
In Fig.\ \ref{normallayer}
we show the self-consistent OP, $\Delta (\bR,T=0.05T_c)$,
for a mean free path of  0.3 coherence lengths in
the shaded region, and of 15 coherence lengths elsewhere. 
The bulk OP
is of the form $k_x+ik_y$, and near the surface a secondary OP
component, $k_x-ik_y$, is induced. As can be seen from Fig.\
\ref{normallayer}, both components are suppressed in the surface layer
where scattering is enhanced,
leading effectively to a normal-conducting layer near the interface.

The presence of a normal-state layer affects strongly point-contact
and tunneling spectra. However, as we show in the insets of Fig.\
\ref{normallayer}, the form of the spectra in the presence of a
normal-state layer is in agreement with experiment (c.f. inset in
Fig.\ \ref{IexcofB} and Ref.\ \onlinecite{LGL+00}). The excess current
is reduced by such a surface layer (see lower inset in Fig.\
\ref{normallayer}). Also, the tunneling  spectra show a pronounced
zero-energy anomaly in contrast to the clean surface. The temperature
dependence of the excess current near $T_c$ is only weakly affected by
a normal-conducting surface layer, leaving the results discussed above
unaltered. 

{\it Conclusions.}---We presented point-contact measurements on the
unconventional superconductor Sr$_2$RuO$_4$ as a function of
temperature and applied magnetic field. The excess current exhibits
linear behavior both in temperature and magnetic field over a large
range. Using these findings we derive a scaling relation between 
the excess current and the order parameter [Eq. (\ref{IofD})]. 

We discuss this result within the theory of $p$-wave spin-triplet
superconductivity. 
We find that that the excess current in unconventional superconductors 
is not necessarily proportional to the order parameter.
In order to account for the wide range over which the linear behavior
of the excess current holds experimentally we extend the pure $p$-wave
theory to take into account 
scattering between quasiparticles and
low-energy bosonic fluctuations, probably originating
from spin fluctuations \cite{Millis88}. The extended theory yields a
very good agreement with the measured excess current and yields a scaling
exponent $\nu=1/3$.
Furthermore it can account for the large $\Delta(0)/k_{\rm B} T_{\rm c }$-ratios
obtained from point-contact measurments
\cite{LGL+00}. 
Finally, we show that surface effects should be considered
for a satisfactory reproduction of the point-contact spectra. 

In closing, we mention that a recent Ginzburg-Landau analysis, assuming a
third order phase transition induced by gapless excitations in the
superconducting phase, yields the correct temperature dependence to
account for the data, at least close to $T_{\rm c}$
\cite{Wern03b,Wern03c}. As shown in Ref. \cite{EFF01} the $p$-wave
channel of superconductivity may be only marginally dominant assuming
that pairing in Sr$_2$RuO$_4$ is mediated by incommensurate spin
fluctuations. In this case 
the presence of fluctuations in the OP is not unlikely.

{\it Acknowledgments.}---We thank F.\ Lichtenberg, Y. Maeno, and
Z.~Q. Mao for supplying the samples. FL was supported by the Deutsche
Forschungsgemeinschaft through the Graduiertenkolleg ``Anwendungen der
Supraleitung''. RW was supported by the Center for Functional
Nano\-struc\-tures at Universit\"at Karlsruhe.

\end{document}